\documentclass[12pt,preprint]{aastex62}
\usepackage{color}
\usepackage{graphicx}
\usepackage{longtable}
\usepackage{threeparttable}
\usepackage{url}
\usepackage{subfigure}
\usepackage{amsmath}
\usepackage{lipsum}
\usepackage{txfonts}
\usepackage{natbib}
\usepackage{ucs}
\shorttitle{RM variations and reversals of FRBs from massive binary systems}
\shortauthors{Zhao et al.}

\begin{document}
\title{Rotation Measure Variations and Reversals of Repeating FRBs in Massive Binary Systems}

\author[0000-0002-2171-9861]{Z. Y. Zhao}
\affiliation{School of Astronomy and Space Science, Nanjing University, Nanjing 210093, China}

\author[0000-0001-6545-4802]{G. Q. Zhang}
\affiliation{School of Astronomy and Space Science, Nanjing University, Nanjing 210093, China}

\author[0000-0003-4157-7714]{F. Y. Wang}
\affiliation{School of Astronomy and Space Science, Nanjing University, Nanjing 210093, China}
\affiliation{Key Laboratory of Modern Astronomy and Astrophysics (Nanjing University), Ministry of Education, Nanjing 210093, China}

\author[0000-0002-7835-8585]{Z. G. Dai}
\affiliation{Department of Astronomy, School of Physical Sciences, University of Science and Technology of China, Hefei 230026, Anhui, China}
%\affiliation{School of Astronomy and Space Science, Nanjing University, Nanjing 210093, China}

\correspondingauthor{F. Y. Wang}
\email{fayinwang@nju.edu.cn}

\begin{abstract}
Recent observations discovered that some repeating fast radio bursts (FRBs) show complicated variations and reversals of Faraday rotation measures (RMs), indicating that the sources of these FRBs are embedded in a dynamically magnetized environment. One possible scenario is that repeating FRBs are generated by pulsars in binary systems, especially containing a high-mass companion with strong stellar outflows. Here, we study the RM variations caused by stellar winds, and a possible stellar disc. If the magnetic field is radial in the stellar wind, RM will not reverse except if the magnetic axis inclination angle is close to $90 ^\circ$. For the toroidal magnetic field in the wind, RM will reverse at the super conjunction. For the case of the toroidal field in the disc, the RM variations may have a multimodal and multiple reversal profile because the radio signals travel through different
components of the disc during periastron passage. We also apply this model to FRB 20180916B. By assuming its 16.35-day period is from a slowly rotating or freely precessing magnetar, we find that the secular RM variation can
be explained by the
periastron passage of a magnetar in a massive binary system. In addition, the clumps in the stellar wind and disc can cause short time-scale ($< 1$ day) variations or reversals of RM. Therefore, long-term monitoring of RM variations can reveal the environments of repeating FRBs.
\end{abstract}


\section{Introduction}
Fast radio bursts (FRBs) are extragalactic millisecond-duration radio transients with high brightness temperatures and large dispersion measures (DMs). FRBs were first discovered more than a decade ago by \cite{Lorimer2007} and we have known little about their physical origins (see reviews of \citealt{ZB2020,Xiao2021,Petroff2022}). At least some FRBs originate from magnetars since FRB 200428 was detected from the Galactic magnetar SGR J1935+2154 \citep{Bochenek2020,CHIME2020}. The magnetar or pulsar related models are proposed to interpret the origins of FRBs \citep{Murase2016,Beloborodov2017,Kashiyama2017,Metzger2017,Wang2017,Yang2018,Metzger2019,Lu2020,Wang2020}. However, for the repeating FRBs being active for several years, the rotation energy is insufficient. The pulsar should be highly magnetized and FRBs shoud be powered by the magnetic energy \citep{Lyutikov2017}.

Except for the very few FRBs close to us, we cannot make detailed observations of their surrounding medium, which is crucial for diagnosing their origin. For cosmological FRBs, we can indirectly obtain information about their local environment or progenitors through their polarization, DM, Faraday rotation measure (RM), scintillation, and scattering. The differences in observed DMs and RMs of FRBs mean that there may be different origins or that they are all from magnetars but in different environments. The large ($\sim10^{5}$ rad m$^{-2}$) and decreasing RM of FRB 20121102A \citep{Michilli2018,Hilmarsson2021} may be from the magnetar wind nebula \citep{Margalit2018,Zhao2021a}, a pulsar near a supermassive black hole (SMBH, \citealt{Zhang2018,Katz2021,Yang2022}) or a hyper-accreting X-ray binary \citep{Sridhar2022}. The extreme DM of FRB 20190520B may be related to the host galaxy \citep{Niu2022} or a young supernova remnant (SNR, \citealt{Zhao2021b,Katz2022}).

The observed RM, especially the extremely large or evolving values, can better reflect the properties of the magnetized environment around the source. Recently, the RM variations or reversals have been observed for some FRBs, such as FRB 20201124A \citep{Xu2022} and FRB 20190520B \citep{Anna-Thomas2022,Dai2022}. A similar RM evolution has been found through the periastron passage of PSR B1259$-$63/LS 2883 (e.g., \citealt{Johnston1996,Johnston2005}). \cite{Wang2022} has proposed that FRB 20201124A and FRB 20190520B may reside in a magnetar/Be star binary system. The fast-rotating Be star will form a decretion disc. RM is expected to be variable when the bursts interact with the disc. The RM reversals of FRB 20201124A and FRB 20190520B can be also naturally explained in this model.

More recently, the RM variations of FRB 20180916B have been reported \citep{Mckinven2022}. FRB 20180916B was first discovered by the Canadian Hydrogen Intensity Mapping Experiment (CHIME) telescope with a low RM of $-114.6 \pm 0.6$ rad m$^{-2}$ \citep{CHIME/FRBCollaboration2019}. It was localized in a nearby massive spiral galaxy at $z=0.0337 \pm 0.0002$ \citep{Marcote2020}. The following monitoring showed that the DM has very little variation \citep{Nimmo2021,Pastor-Marazuela2021,Mckinven2022}. The $\left|\mathrm{RM}\right|$ has a small and stochastic variation before 2021 April \citep{Chawla2020,Pleunis2021,Mckinven2022}, and has a secular decrease during 2021 April and 2021 December \citep{Mckinven2022}.

Besides the above secular DM and RM variations, short-time RM variations have been measured in a pulsar binary system PSR B1744$-$24A \citep{Li2022}. A turbulent medium or cloud in SNR has been proposed to explain the irregular RM variations \citep{Yang2022}. There may be an inhomogeneous medium near the FRB source,
and the relative motion between the source and the medium causes
stochastic RM variations. From observations, stellar winds \citep{Moffat1994,Puls2006} or stellar discs \citep{Chernyakova2021} are found to be clumpy (e.g., the short variability on keV/GeV light curve, see \citealt{Apparao1991,Chernyakova2020,Chernyakova2021}). Here, we will estimate the stochastic RM variation from the clumpy stellar wind and disc.

In this work, we study the effects of stellar winds and discs on the observational properties of FRBs in a high mass X-ray binary (HMXB) or a high mass $\gamma-$ray binary (HMGB) (see Fig. \ref{fig}). The motivations to consider FRBs originating in binary systems are as follows. First, \cite{Tendulkar2021} found that FRB 20180916B is 250 pc away from the nearest region of active star formation in its host galaxy. The inferred age of the source should be about a few million years to traverse the observed separation for the typical kick velocity of a pulsar, which is opposed to a young active magnetar born in core-collapse supernova. However, the separation and the inferred age can be compatible with a highly magnetised neutron star in an HMXB or an HMGB. Second, similar RM variations and reversals of repeating FRBs have been observed both in repeating FRBs \citep{Xu2022,Anna-Thomas2022,Dai2022} and a binary system containing a pulsar and a massive Be star (PSR B1259$-$63/LS 2883) \citep{Johnston1996,Johnston2005}. \cite{Wang2022} found that RM variations and reversals can be well understood in a magnetar/Be star binary system. Third, short-time RM variations have been measured in a pulsar binary system PSR B1744$-$24A \citep{Li2022}. Massive stars usually have strong stellar winds, which is not discussed in detail in previous works \citep{Wang2022,Yang2022}. More importantly, the interactions between the stellar wind and the pulsar wind will form a bow shock cavity, which will play an essential role in the orbital modulations of RM, DM, and the free-free absorption process. In addition to a polar wind, a decretion disc will be formed for some Oe/Be stars. However, the disc will be truncated due to the tidal torques from the pulsar \citep{Reig2011}. Luckily, the tidal interactions can be neglected in an eccentric orbital \citep{Okazaki2011}. We will only discuss the DM and RM from the disc with a high eccentricity orbital for simplicity.

This letter is organized as follows. In Section \ref{shock}, the wind interaction and the geometric description of the binary system are shown. Taking FRB 20180916B as an example, Section \ref{dmrm} gives the results of RM, DM, and free-free absorption process from the stellar wind and disc. The stochastic DM and RM variations caused by the clumps in the stellar wind and disc are given in Section \ref{sec:clumps}. The DM and RM variations for other repeating FRBs and the possible origins of the persistent radio sources (PRSs) in a binary system are discussed briefly in Section \ref{discussion}. Finally, a summary is given in Section \ref{conclusion}.

\section{Wind interactions}\label{shock}
\subsection{The stellar outflows}
For a massive star, the mass-loss is important via an approximately isotropic stellar wind. The electron density of the stellar wind from the radial distance $r$ is given by
\begin{equation}
    n_{\mathrm{w}}(r)=n_{\mathrm{w}, 0}\left(\frac{r}{R_{\star}}\right)^{-2},
\end{equation}
where $n_{\mathrm{w}, 0}$ is the number density at the stellar surface and $R_{\star}$ is the radius of the star. The constant mass-loss rate $\dot{M}$ and the assumption of the completely ionized wind is used in this work. In this case, $n_{\mathrm{w}, 0}$ can be written as $n_{\mathrm{w}, 0}=\dot{M} / 4 \pi R_{\star}^{2} v_{\mathrm{w}}(r) \mu_{\mathrm{i}} m_{\mathrm{p}}$, where $v_{\mathrm{w}}(r)=v_{\mathrm{w},\infty}(1-R_{\star}/r)$ is the wind velocity with $v_{\mathrm{w},\infty}$ being the terminal speed of the wind, $\mu_{\mathrm{i}}$ is the mean ion molecular weight and $m_{\mathrm{p}}$ is the mass of protons. At a large distance ($r\gg R_{\star}$), the wind velocity is $v_{\mathrm{w}}\simeq v_{\mathrm{w},\infty}$. The typical value of mass-loss rate and wind velocity for Be stars is found to be $10^{-11}-10^{-8} M_{\odot}~ \mathrm{yr}^{-1}$ and $v_{w}\sim 10^{8} \mathrm{~cm} \mathrm{~s}^{-1}$ from observations \citep{Snow1981,Krticka2014}. For typical stellar outflows with hydrogen abundance $X\sim0.7$ \citep{Dubus2013}, $\mu_{\mathrm{i}}=4 /(1+3 X) \simeq 1.29$. For some Oe/Be stars, the slow equatorial outflow will form a decretion disc. The disc usually has a vertical Gaussian density profile \citep{Carciofi2006}
\begin{equation}
    n_{\mathrm{d}}\left(r_{\mathrm{d}}, z\right)=n_{\mathrm{d}, 0}\left(\frac{r_{\mathrm{d}}}{R_{\star}}\right)^{-\beta_{\mathrm{d}}} \exp \left[-\frac{z_{\mathrm{d}}^{2}}{2 H^{2}\left(r_{\mathrm{d}}\right)}\right],
\end{equation}
where $r_{\mathrm{d}}$ and $z_{\mathrm{d}}$ are the radial and vertical distances, respectively. $n_{\mathrm{d}, 0}=\rho_{\mathrm{d}, 0}/\mu m_{\mathrm{p}}$ is the disc base number density at $R_{\star}$, where $\mu\sim0.62$ is the molecular weight in the disc and the density of the disk near stellar surface $\rho_{\mathrm{d}, 0}$ is between 10$^{-13}$ g cm$^{-3}$ and 10$^{-10}$ g cm$^{-3}$ \citep{Rivinius2013}. $\beta_{\mathrm{d}}=2-4$ is the disc density slope \citep{Rivinius2013}. The scale height of the disc is
\begin{equation}
    H(r_{\mathrm{d}})=c_{s}\left(\frac{r_{\mathrm{d}}}{G M_{\star}}\right)^{1 / 2} r_{\mathrm{d}},
\end{equation}
where $M_{\star}$ is the star mass and $c_{s}=\left(k T_\star / \mu m_{\mathrm{p}}\right)^{1 / 2}$ is the isothermal sound speed.

\begin{figure}
    \centering
    \includegraphics[width = 0.8\textwidth]{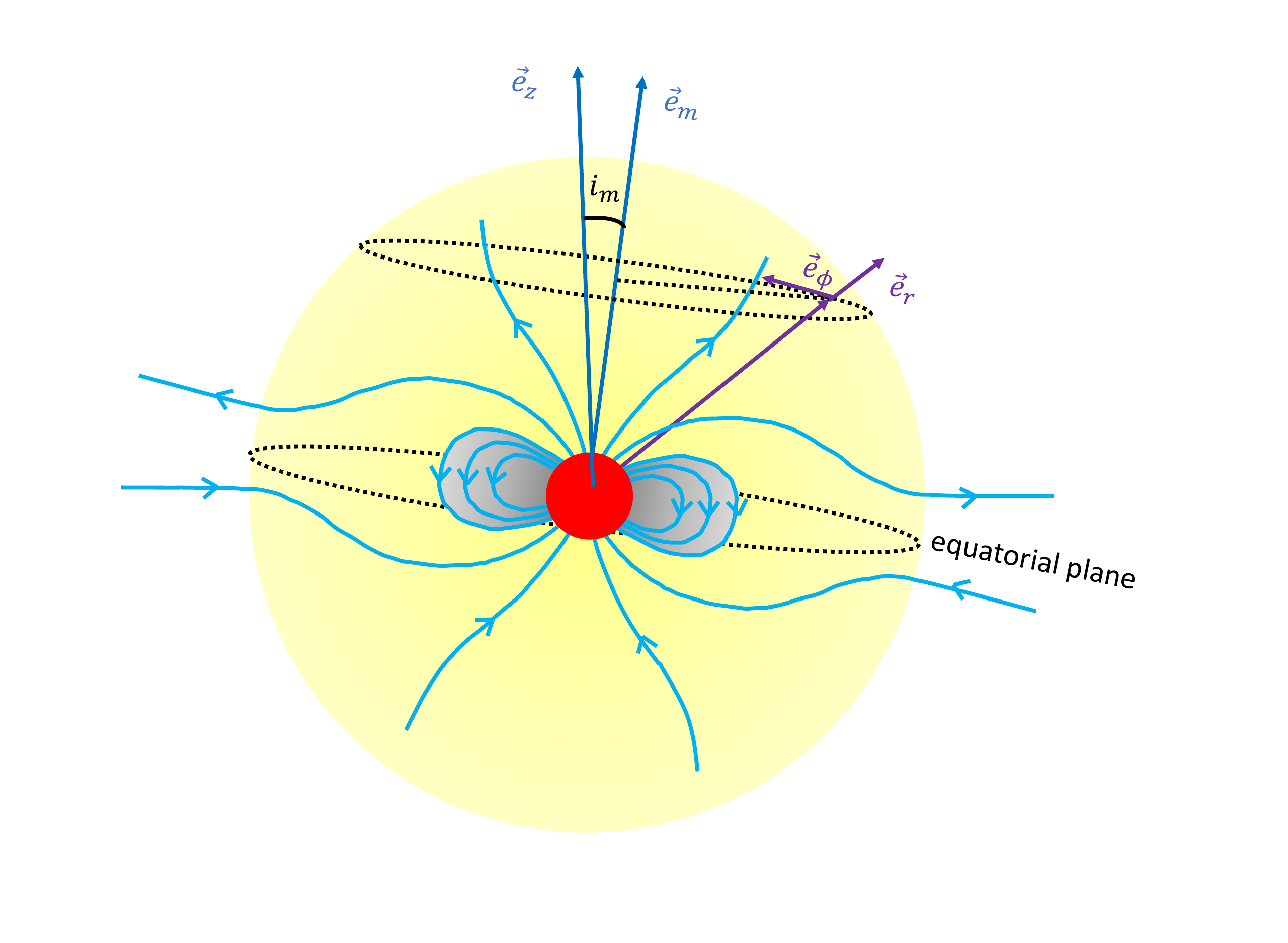}
    \caption{A schematic diagram of the magnetic field of a massive star. The orbital angular momentum is in the $z$-direction and the magnetic axis inclination angle is $i_{\mathrm{m}}$. The magnetic field of the star is dipolar within the Alfvén radius $R_{\mathrm{A}}$ (shaded region). The magnetic field will become radial at a large distance because of the drag effect of the stellar wind and the fast rotation of OB stars will make the magnetic field become toroidal \citep{Usov1992}.}
    \label{mag}
\end{figure}

The orbital angular momentum is in the $z$-direction and the magnetic axis inclination angle is $i_{\mathrm{m}}$ (see Fig. \ref{mag}). The magnetic field of the star is dipole ($B \propto r^{-3}$) within the Alfvén radius $R_{\mathrm{A}}$. The magnetic field will become radial at a large distance because of the drag effect of the stellar wind and the fast rotation of OB stars will make the magnetic field become toroidal \citep{Usov1992}. Thus, for a massive star, the magnetic field in the stellar wind or disc can be described as \citep{Melatos1995,Yang2022}
\begin{equation}
    B_{\mathrm{s}}(r) \sim \begin{cases}B_0\left(\frac{r}{R_{\star}}\right)^{-3}, & R_{\star}<r<R_{\text{A}}, \\ B_0\left(\frac{R_{\text{A}}}{R_{\star}}\right)^{-3}\left(\frac{r}{R_{\text{A}}}\right)^{-\alpha}, & r>R_{\text{A}},\end{cases}
\end{equation}
where $B_{0}$ is the magnetic field strength at $R_{\star}$. We have $\alpha=2$ for a radial field and $\alpha=1$ for a toroidal field. he Alfvén radius $R_{\mathrm{A}}$ can be estimated by the balance of the stellar wind's ram pressure and the magnetic pressure
\begin{equation}
    \frac{1}{2} \rho_w\left(R_A\right) v_w^2 \simeq \frac{1}{8 \pi} B\left(R_A\right)^2.
\end{equation}
The RM variations of HMGBs (e.g., PSR B1259$-$63/LS 2883 \citealt{Johnston1996}) can be well understood in the context of toroidal field in stellar disc \citep{Melatos1995}. The temperature of the stellar wind decay with the following power-law because of adiabatic cooling \citep{Kochanek1993,Bogomazov2005}
\begin{equation}\label{Tw}
    T_{\mathrm{w}}(r)=T_{\star}\left(\frac{r}{R_{\star}}\right)^{-\beta_{\mathrm{w}}},
\end{equation}
where $T_{\star}$ is the surface effective temperature of the star and $\beta_{\mathrm{w}}=2/3$. The temperature profile of the vertically isothermal disc is given by \citep{Carciofi2006}
\begin{equation}\label{Td}
    T_{\mathrm{d}}\left(r_{\mathrm{d}}\right)=\frac{T_{\star}}{\pi^{1 / 4}}\left[\arcsin \left(\frac{R_{\star}}{r_{\mathrm{d}}}\right)-\left(\frac{R_{\star}}{r_{\mathrm{d}}}\right) \sqrt{1-\left(\frac{R_{\star}}{r_{\mathrm{d}}}\right)^{2}}\right]^{1 / 4}.
\end{equation}
The parameters of the massive star we adopted are: the stellar mass $M_{\star}=30~M_{\odot}$, the stellar radius $R_{\star}=10~R_{\odot}$, the surface effective temperature $T_{\star}=2\times10^4$ K and the wind velocity $v_{w}= 3 \times 10^{8} \mathrm{~cm} \mathrm{~s}^{-1}$, which are consistent with the observations (e.g., LS 2883 \citealt{Negueruela2011}; LS 5039 \citealt{Casares2005}).

\subsection{The terminal shock}
\begin{figure}
    \centering
    \includegraphics[width = 1\textwidth]{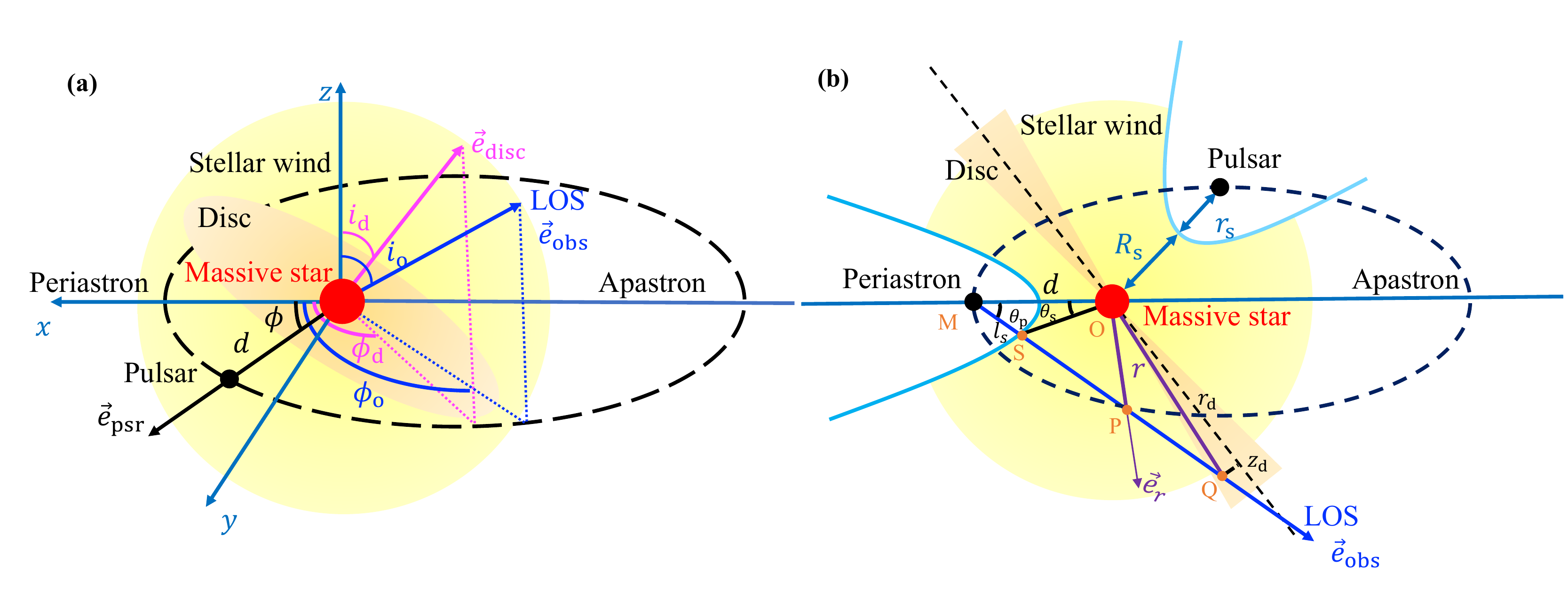}
    \caption{A schematic diagram of the pulsar/massive star binary (panel a, pulsar wind is not included) and the interactions between the smooth stellar outflows and the pulsar wind (panel b).
    The descriptions of parameters are given in Table \ref{tab:my_label}.}
    \label{fig}
\end{figure}

The pulsar wind will be terminated by the stellar wind and the wind interactions will form a bow shock cavity, which is shown in Fig. \ref{fig}(b). The momentum rate ratio of the pulsar and the stellar wind is defined as
\begin{equation}
    \eta=\frac{L_{\mathrm{sd}} / c}{\dot{M} v_{\mathrm{w}}},
\end{equation}
where $L_{\mathrm{sd}}$ is the spin-down luminosity of the pulsar. If $\eta>1$, the pulsar wind is stronger, and the shock bend back to the star. In this case, there is no RM variation for most orbital
phase, which contradicts with the observations of repeating FRBs. Therefore, we consider the case of $\eta<1$. The distances from the termination shock to the pulsar and massive star are
\begin{equation}
    r_{\mathrm{s}}=d \frac{\eta^{1 / 2}}{1+\eta^{1 / 2}},
\end{equation}
and
\begin{equation}
    R_{\mathrm{s}}=d \frac{1}{1+\eta^{1 / 2}},
\end{equation}
respectively. The orbital separation $d$ at phase $\phi$ is
\begin{equation}
    d=\frac{a\left(1-e^{2}\right)}{1+e \cos \phi},
\end{equation}
where $e$ is the eccentricity of the orbit. The semimajor axis $a$ is given by Kepler’s third law
\begin{equation}
    \frac{P_{\mathrm{orb}}^{2}}{a^{3}}=\frac{4 \pi^{2}}{G\left(M_{\star}+m\right)},
\end{equation}
where $P_{\mathrm{orb}}$ is the orbital period, $M_{\star}$ and $m=1.4~M_{\odot}$ are the mass of massive star and pulsar, respectively.

To calculate the DM, RM and the free-free optical depth along the LOS, we adopt the geometric description of the binary system from \cite{Chen2021}. As shown in Fig. \ref{fig}(a), the unit vector of the observer is
\begin{equation}
    \boldsymbol{e}_{\mathrm{obs}}=\left(\sin i_{\mathrm{o}} \cos \phi_{\mathrm{o}}, \sin i_{\mathrm{o}} \sin \phi_{\mathrm{o}}, \cos i_{\mathrm{o}}\right),
\end{equation}
where $i_{\mathrm{o}}$ and $\phi_{\mathrm{o}}$ are the inclination angle and true anomaly of the observer, respectively. The normal unit vector of the disc is
\begin{equation}
    \boldsymbol{e}_{\mathrm{disc}}=\left(\sin i_{\mathrm{d}} \cos \phi_{\mathrm{d}}, \sin i_{\mathrm{d}} \sin \phi_{\mathrm{d}}, \cos i_{\mathrm{d}}\right),
\end{equation}
where $i_{\mathrm{d}}$ and $\phi_{\mathrm{d}}$ are the inclination angle and true anomaly of the disc normal unit vector, respectively. As shown in Fig. \ref{fig}(b), the unit vector in the direction connecting any point along the LOS to the massive star is
\begin{equation}
    \boldsymbol{e}_{\mathrm{r}}=\left(d \cdot \boldsymbol{e}_{\mathrm{psr}}+l \cdot \boldsymbol{e}_{\mathrm{obs}}\right) / r,
\end{equation}
where $\boldsymbol{e}_{\mathrm{psr}}=(\cos \phi, \sin \phi, 0)$ is the unit vector in the direction of the pulsar and $l$ is the distance from the pulsar. The distance from the arbitrary point along LOS to the massive star is
\begin{equation}
    r^{2}=d^{2}+l^{2}+2d l\left(\boldsymbol{e}_{\mathrm{psr}} \cdot \boldsymbol{e}_{\mathrm{obs}}\right).
\end{equation}
If the LOS passes through the stellar disc, the radial and vertical distances in the disc are given by
\begin{equation}
    r_{\mathrm{d}}=r \sin \theta, z_{\mathrm{d}}=r \cos \theta,
\end{equation}
where $\theta=\arccos \left(\boldsymbol{e}_{\text {r}} \cdot \boldsymbol{e}_{\mathrm{disc}}\right)$ is the angle between the arbitrary radial direction from the star and the disc normal direction.

\section{DM and RM variations}\label{dmrm}
The observed DM and RM of a cosmological-origin FRB (with the redshift $z$) are contributed by different parts, including the local environment of FRBs, the host galaxy, the intergalactic medium (IGM) and the Milky Way (MW):
\begin{equation}
    \mathrm{DM}_{\mathrm{obs}}=\mathrm{DM}_{\mathrm{MW}}+\mathrm{DM}_{\mathrm{IGM}}+\frac{\mathrm{DM}_{\text {host}}+\mathrm{DM}_{\text {source }}}{1+{z}},
\end{equation}
\begin{equation}
    \mathrm{RM}_{\mathrm{obs}}=\mathrm{RM}_{\mathrm{MW}}+\mathrm{RM}_{\mathrm{IGM}}+\frac{\mathrm{RM}_{\text {host }}+\mathrm{RM}_{\text {source }}}{(1+z)^{2}}.
\end{equation}
Along the paths of FRBs, the medium in MW, IGM and host galaxies are stable on a relatively short time scale. Therefore, the DM and RM variations mainly come from the local plasma of sources, including the contributions from the pulsar wind, the terminal shock shell and the stellar outflows. The electron number density in the pulsar wind can be negligible compared to the stellar wind of massive stars ($n_{\mathrm{BE}} \sim 10^{7} n_{\mathrm{PS}}$, see \citealt{Melatos1995}). What's more, the DM and RM from relativistic electrons should be reduced by a factor of $\gamma^2$, with $\gamma$ being the Lorentz factor of electrons (\citealt{Quataert2000}). Thus, the RM contributed by a pulsar wind is very small and can be ignored.

The thickness $h$ of the terminal shocked shell at the apex is $h=\frac{1}{8}\min(r_{\mathrm{s}}, R_{\mathrm{s}})$ \citep{Luo1990}. The shocked material will lose its energies via adiabatic or radiative cooling processes. The adiabatic dominated shock radiation is consistent with the photon index of keV X-ray and TeV $\gamma$-ray spectrum form PSR B1259$-$63/LS 2883 \citep{Chen2019}. In this case, the asymptotic opening angle between the reverse shock and the contact surface is $\Delta \psi \sim 0$ \citep{Luo1990}. Different from the extended SNR shock shell \citep{Yang2017,Piro2018,Zhao2021a,Zhao2021b}, the DM and RM from the terminal shock shell are also unimportant.

The DM and RM variations from the stellar wind or disc are
\begin{equation}
\begin{aligned}
    &\Delta \mathrm{DM}=\int_{\operatorname{LOS}} n_{\mathrm{e}} \mathrm{d} l, \\
    &\Delta \mathrm{RM}=8.1 \times 10^{5} \int_{\operatorname{LOS}} n_{\mathrm{e}} \boldsymbol{B} \cdot \mathrm{d} \boldsymbol{l},
\end{aligned}
\end{equation}
where the electron number density $n_{\mathrm{e}}$ (in units of cm$^{-3}$), the magnetic field $\boldsymbol{B}$ (in units of G) and the distance $l$ (in units of pc) are given in Section \ref{shock}. The free-free absorption coefficient is \citep{Rybicki1979}
\begin{equation}
    \alpha_{\nu}^{\mathrm{ff}}=0.018 T^{-3 / 2} z_{\mathrm{i}}^{2} n_{\mathrm{e}} n_{\mathrm{i}} \nu^{-2} \bar{g}_{\mathrm{ff}},
\end{equation}
where the Gaunt factor $\bar{g}_{\mathrm{ff}}\sim1$ and the atomic number of the ion $z_{\mathrm{i}}\sim1$ are used in this work. The free–free optical depth of the stellar outflow is
\begin{equation}
    \tau(\nu)=\int_{\mathrm{LOS}} \alpha(\nu) \mathrm{d} l.
\end{equation}
Because we neglect the contributions from the pulsar wind and the terminal shock shell, we can get the DM, RM and $\tau(\nu)$ integrating from the joining point of the LOS and the shock contact discontinuity (CD) surface to infinity numerically. As shown in Fig. \ref{fig}(b), points S, P and Q are the intersection points of LOS and the shock CD surface, arbitrary radial direction of stellar wind and disc, respectively. The pulsar and massive star are regarded as mass points, denoted as points M and O, respectively. The distance from the bow shock CD surface to the pulsar is \citep{Canto1996}
\begin{equation}
    l_{\mathrm{s}}=d \sin \theta_{\mathrm{s}} \csc \left(\theta_{\mathrm{s}}+\theta_{\mathrm{p}}\right),
\end{equation}
where $\theta_{\mathrm{s}}$ and $\theta_{\mathrm{p}}$ are the angles between the line connecting the two stars and the line SM and SO, respectively. For most HMXBs, the pulsar wind is much weaker than that of stars ($\eta \ll 1$). In this way, the angle $\theta_{\mathrm{s}}$ is approximated as follows \citep{Canto1996}
\begin{equation}
    \theta_{\mathrm{s}}=\left\{\frac{15}{2}\left[\sqrt{1+\frac{4}{5} \eta\left(1-\theta_{\mathrm{p}} \cot \theta_{\mathrm{p}}\right)}-1\right]\right\}^{1 / 2},
\end{equation}
where
\begin{equation}
    \theta_{\mathrm{p}}=\pi-\arccos \left(\boldsymbol{e}_{\mathrm{psr}} \cdot \boldsymbol{e}_{\mathrm{obs}}\right)
\end{equation}
depends on the orbital phase.

\subsection{DM and RM from the stellar wind: application to FRB 20180916B}\label{wind}

A periodic ($\sim16.35$ d) bursting activity was found for FRB 20180916B \citep{Chime/FrbCollaboration2020}, which may originate from orbital motion \citep{Dai2020,Ioka2020,Lyutikov2020,Gu2020,Li2021,Wada2021}, precession \citep{Levin2020,Tong2020,Yang2020,Zanazzi2020,Katz2021b,Sridhar2021,LDZ2021,Wei2022,Chen2022} or ultralong spin of magnetars \citep{Beniamini2020}. The chromaticity was first found by \cite{Pleunis2021} and \cite{Pastor-Marazuela2021}. Very recently, \cite{Bethapudi2022} extented the chromatic activity window of FRB 20180916B to higher frequencies. The chromaticity is consistent with binary models described in \cite{Wada2021} or slowly rotating/freely precessing models described in \cite{LDZ2021}. However, the RM variations are unrelated to the $\sim16.35$ d period \citep{Mckinven2022}, which is inconsistent with the quasi-periodic variations of a binary system.
Thus, the assumption of the 16.35-day period being from a precessing (or rotating) magnetar is used in this work. The burst is produced by the coherent curvature radiation in the magnetar's magnetosphere \citep{Yang2018,Lu2020} and the chromaticity is caused by the altitude-dependent radio emission \citep{LDZ2021}. In addition, the clustered polarization position angles (PPAs) in phase have been found \citep{Bethapudi2022}, which has been predicted in \cite{LDZ2021}.

Motivated by the observations that the RM variation is unrelated to the 16.35-day period \citep{Mckinven2022}, we consider that FRB 20180916B is in a binary with a long orbital period (see Fig. \ref{fig}). The secular RM variation may be
caused by the periastron or superior conjunction (SUPC) passage of the pulsar in an HMXB or an HMGB. In this way, the current observations of FRB 20180916B can be explained simultaneously.

\begin{figure}
    \centering
    \includegraphics[width = 0.8\textwidth]{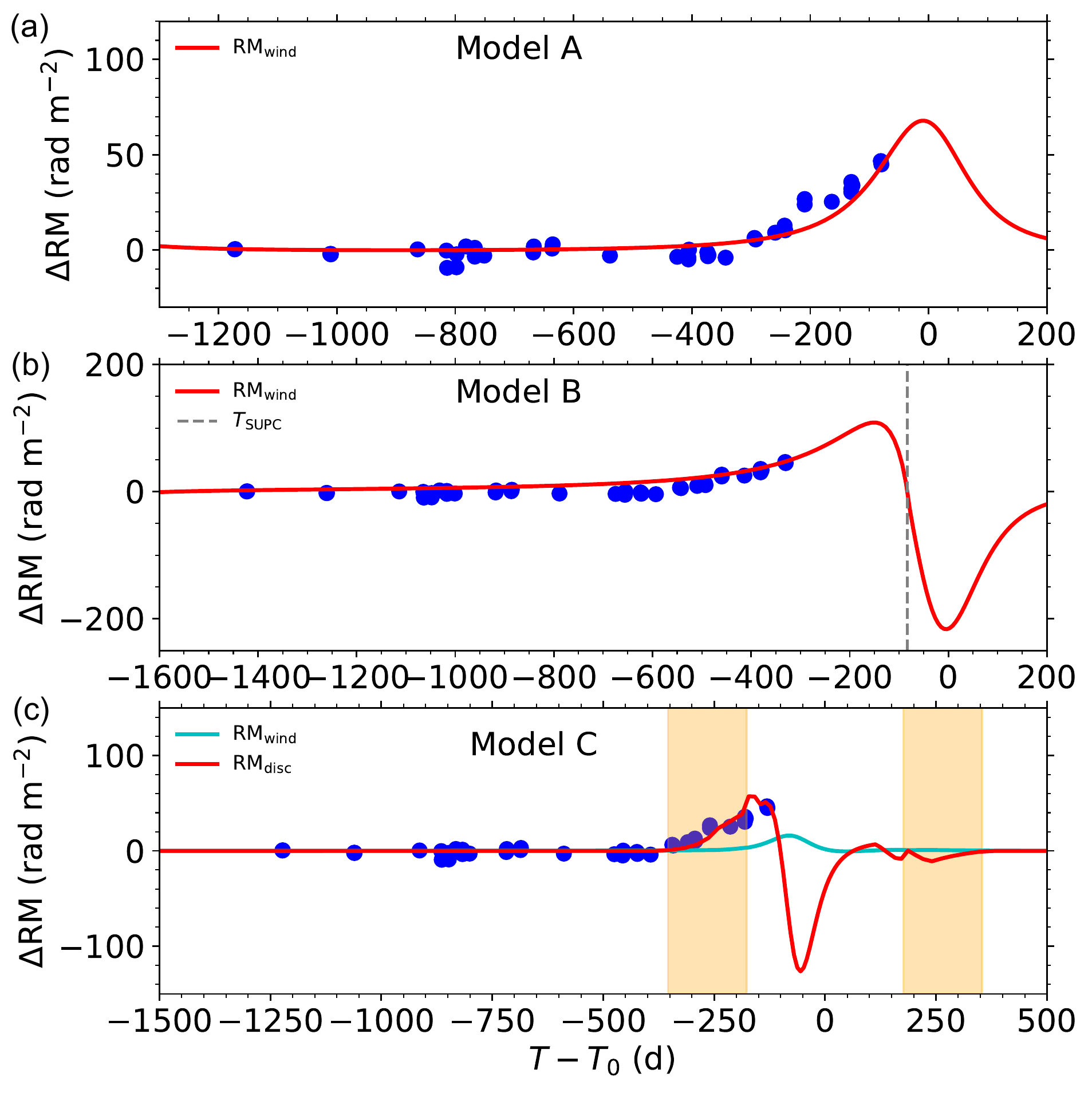}
    \caption{The RM variations from the stellar wind and disc. The blue circles represent the observed RM variations $\Delta \mathrm{RM_{obs}}=\mathrm{RM_{obs}}-\mathrm{RM_{obs,0}}$ from \cite{Mckinven2022}, and $\mathrm{RM_{obs,0}}=-114$ rad m$^{-2}$ is the value of the non-evolving phase \citep{Chime/FrbCollaboration2020,Nimmo2021,Pastor-Marazuela2021,Mckinven2022}. The fitting results are shown in solid curves, and the model parameters are listed in Table \ref{parameters}. (a): The fitting result of the radial magnetic field in the wind. Near periastron, RM variations reach the peak. (b): The fitting result of the toroidal wind magnetic field. RM will reverse before and after the SUPC ($T_{\mathrm{SUPC}}=-83.4$ d). (c): The fitting result of the disc model and the contributions of the stellar wind is also considered. The toroidal magnetic field in the disc is used and the magnetic field in the stellar wind is assumed to be radial. The $\Delta$RM from the disc and stellar wind is shown in red and cyan curves, respectively. The pulsar will pass through the inclined disc twice, i.e., before and after periastron. The orange-shaded regions represent the time when the pulsar passes through the disc. The pressure from the disc will reduce the momentum rate ratio of the pulsar and the stellar wind ($\sim0.01\eta$). The radio signals travel through different components of the disc when a pulsar passes through the disc near periastron, whcih leads to complicated RM variations (e.g., with some minor structures and multiple reversals).}
    \label{rm}
\end{figure}

Below, we fit the RM variation of FRB 20180916B. The typical values we adopted are listed in Table \ref{tab:my_label}. To explain nearly non-evolved RM from the year 2018 to 2021 and the secular decrease of $|\mathrm{RM}|$ from 2021 April to 2021 December \citep{Mckinven2022}, the orbital period must be longer than 3 yr. It must be noted that this orbital period is not related to the orbiting models producing the periodic activity of this source. The fitting results for the cases of $P=1600$ d, $e=0.5$ (Model A) and $P=2000$ d, $e=0.5$ (Model B) are shown in Fig. \ref{rm} and the parameters are listed in Table \ref{parameters}. For the case of the stellar wind, the magnetic axis inclination angle is assumed to be $\sim0^\circ$ and the discussion of different magnetic axis inclination angles is given in Section \ref{190520}. The blue circles represent the observed RM variations $\Delta \mathrm{RM_{obs}}=\mathrm{RM_{obs}}-\mathrm{RM_{obs,0}}$ from \cite{Mckinven2022}, and $\mathrm{RM_{obs,0}}=-114$ rad m$^{-2}$ is the value of the nearly non-evolving phase \citep{Chime/FrbCollaboration2020,Nimmo2021,Pastor-Marazuela2021,Mckinven2022}. RM variations are shown in red curves, defined as $\Delta \mathrm{RM}=\mathrm{RM}(t)-\mathrm{RM_{0}}$, where $\mathrm{RM_{0}}$ is the minimum value of the orbital modulated RM occurring around apastron or inferior conjunction (INFC).

For Model A, the magnetic field of stellar wind is assumed to be radial ($\boldsymbol{B}_{\mathrm{s}}=B_{0}\left(R_{\text{A}}/R_{\star}\right)^{-3}\left(r/R_{\text{A}}\right)^{-2}\boldsymbol{e}_{\mathrm{r}}$). The spin-down luminosity is $L_{\mathrm{sd}}=10^{36}$ erg s$^{-1}$, which is consistent with the observations of the pulsars in Be/X-ray binaries (e.g., $L_{\mathrm{sd}}=8\times10^{35}$ erg s$^{-1}$ for PSR B1259$-$63 \citealt{Manchester1995}). The mass-loss rate and the stellar surface magnetic field is $10^{-8} M_{\odot}~ \mathrm{yr}^{-1}$ and 5 G, respectively. The periastron time $T_0$ is on MJD 59650. Near periastron, RM variations reach the peak. For radial magnetic field in the wind, RM will not change sign before and after periastron. The toroidal magnetic field ($\boldsymbol{B}_{\mathrm{s}}=B_{0}\left(R_{\text{A}}/R_{\star}\right)^{-3}\left(r/R_{\text{A}}\right)^{-1}\boldsymbol{e}_{\phi}$) in the stellar wind is used in Model B. The spin-down luminosity is $L_{\mathrm{sd}}=2\times10^{35}$ erg s$^{-1}$, which is similar as that of PSR J2032$+$4127 \citep{Camilo2009}. The mass-loss rate, the stellar surface magnetic field and the periastron time are $10^{-9} M_{\odot}~ \mathrm{yr}^{-1}$, 2 G and MJD 59900, respectively. For the case of the toroidal magnetic field, RM will reverse before and after the SUPC ($T_{\mathrm{SUPC}}=-83.4$ d).

\begin{figure}
    \centering
    \includegraphics[width = 1\textwidth]{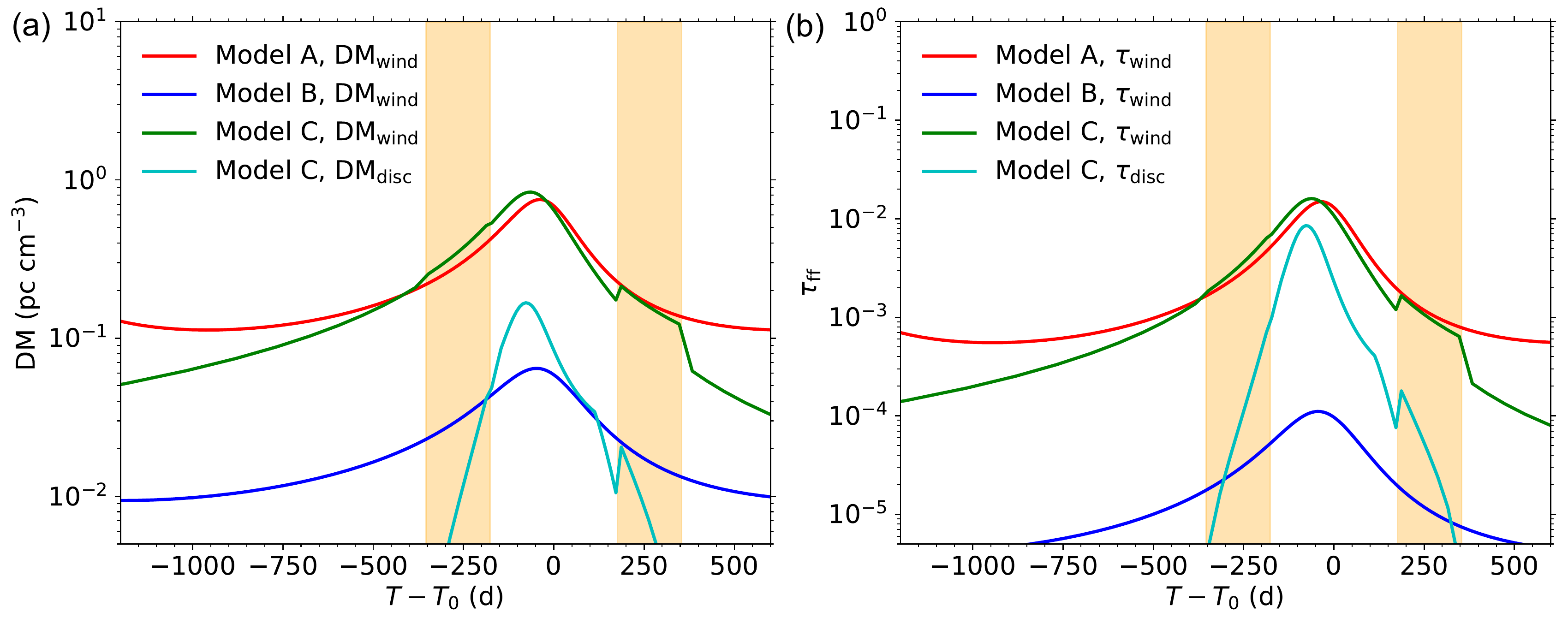}
    \caption{The DM and the free-free optical depth contributed by the stellar wind and disc. The model parameters are listed in Table \ref{parameters}. The peak DM and $\tau_{\mathrm{ff}}$ from the isotropic stellar wind is reached near periastron (see red, blue and green curves). The pulsar will pass through the inclined disc twice before and after periastron. The time when the pulsar passes through the disc is shown in the orange-shaded regions. The dispersion or free-free absorb process from the disc only occurs when the pulsar moves close to the disc. The pressure from the disc will reduce the momentum rate ratio of the pulsar and the stellar wind ($\sim0.01\eta$) and cause small variations on DM and $\tau_{\mathrm{ff}}$ (see green and cyan curves).}
    \label{dm}
\end{figure}

The radio signals from HMXBs with short periods will be obscured due to the free-free absorption of stellar outflows, such as LS I +61$^{\ensuremath{\circ}}$303 (see the study of \citealt{Zdziarski2010}). Nonetheless, we can still receive the radio signals when the pulsar wind cavity\footnote{For the short-period system, the pulsar wind cavity will become a spiral-shape (or a funnel-shape used in \citealt{Ioka2020}) due to the strong Coriolis shock \citep{Bosch-Ramon2011,Bosch-Ramon2015}. However, the effect caused by Coriolis force can be ignored for the long-period system \citep{Chen2021}.} points toward us, which has been proposed to explain the `active window' of periodic FRBs \citep{Ioka2020,Lyutikov2020,Wada2021}. However, the RM variations are unrelated to the 16.35-day period \citep{Mckinven2022}. Here, we consider that the RM variations are caused by the long orbital period. The DM contributions and the free-free optical depth of the stellar wind are shown in Fig. \ref{dm}. The DM variations and the free-free absorption process can be neglected, which is consistent with observations \citep{Chawla2020, Chime/FrbCollaboration2020,Nimmo2021,Pastor-Marazuela2021,Pleunis2021,Mckinven2022}.

\subsection{DM and RM from the stellar disc: application to FRB 20180916B}\label{disc}
Some massive stars, i.e., Be stars, have a decretion disk, which is considered the origin of the emission lines and infrared excess \citep{Reig2011,Rivinius2013}. The orbital plane and the disc are usually misaligned in a Be/X-ray binary due to supernova kicks of pulsars \citep{Brandt1995,Martin2009}. However, the disc inclination angle is still unknown. From the observations, the disc inclination angle is found to be from 25$^{\circ}$ to 70$^{\circ}$ for PSR B1259$-$63/LS 2883 \citep{Martin2009}. By fitting to the DM and RM variations of PSR B1259$-$63, the disc inclination angle is found to be from 10$^{\circ}$ to 40$^{\circ}$ \citep{Melatos1995}. However, a large inclination angle $i_{\mathrm{d}}=45^\circ$, $i_{\mathrm{d}}=60^\circ$ or $i_{\mathrm{d}}=70^\circ$ is also possible given by the study of radio absorption \citep{Chen2021}, multi-wavelength emissions \citep{Chen2019} or XMM–Newton observations \citep{Chernyakova2006} from PSR B1259$-$63/LS 2883. In this work, the disc inclination angle  $i_{\mathrm{d}}=60^{\circ}$ is used. The magnetic axis direction is assumed the same as that of the disc normal vector. The disc will be truncated due to the tidal torques from the pulsar \citep{Reig2011}, especially in circular orbits \citep{Okazaki2011}. Here, we will discuss the case of the Be/X-ray binaries with high orbital eccentricity to avoid disc truncation. The true anomaly angle of the inclined equatorial disc projected on the orbital plane is \citep{Chen2019,Chen2021}
\begin{equation}
    \Delta \phi_{\mathrm{d}}=\sin ^{-1}\left(\frac{\sin \Delta \theta_{\mathrm{d}}}{\sin i_{\mathrm{d}}}\right),
\end{equation}
where $\Delta \theta$ is the half-opening angle of the disc. The pulsar will pass through the inclined disc twice before and after periastron ($\phi \in\left[\phi_{\mathrm{d}}-\pi/2-\Delta \phi_{\mathrm{d}},\phi_{\mathrm{d}}-\pi/2+\Delta \phi_{\mathrm{d}} \right]$ and $\phi \in\left[\phi_{\mathrm{d}}+\pi/2-\Delta \phi_{\mathrm{d}},\phi_{\mathrm{d}}+\pi/2+\Delta \phi_{\mathrm{d}} \right]$). When the pulsar passes through the disc, the pressure from the disc will further push the shock surface towards the pulsar and make the momentum rate ratio $\eta$ reduced. From the estimation of \cite{Chen2019}, the additional pressure from the disc makes $\sim0.01\eta$.

The fitting result of RM variations from the stellar wind and disc (Model C) is shown in panel (c) of Fig. \ref{rm}. The spin-down luminosity is $L_{\mathrm{sd}}=10^{36}$ erg s$^{-1}$. The disc density slope $\beta_{\mathrm{d}}=3.5$ is used \citep{Carciofi2006,Rivinius2013}. The parameters of the stellar wind and disc models are listed in Table \ref{parameters}. The period and the orbital eccentricity of the binary system are $P=16000$ d and $e=0.86$, respectively. The orbital parameters are similar to the Be/X-ray binary PSR J2032$+$4127/MT91 213 ($P=16000-17000$ d, $e\sim0.9$) estimated by \cite{Ho2017}. Follow the study of RM variations of PSR B1259$-$63, the toroidal magnetic field ($\boldsymbol{B}_{\mathrm{s}}=B_{0}\left(R_{\text{A}}/R_{\star}\right)^{-3}\left(r/R_{\text{A}}\right)^{-1}\boldsymbol{e}_{\phi}$) in the disc is used and the magnetic field in the stellar wind is assumed to be radial ($\boldsymbol{B}_{\mathrm{s}}=B_{0}\left(R_{\text{A}}/R_{\star}\right)^{-3}\left(r/R_{\text{A}}\right)^{-2}\boldsymbol{e}_{\mathrm{r}}$) \citep{Melatos1995}. The mass-loss rate and the stellar surface magnetic field is $10^{-8} M_{\odot}~ \mathrm{yr}^{-1}$ and 1 G, respectively. The half-opening angle of the disc is $\Delta \theta=15^{\circ}$ \citep{Waters1986}. In this case, the RM is dominated by the disc. The periastron time is on MJD 59700. The pulsar will pass through the inclined disc twice before and after periastron. The orange-shaded regions represent the time when the pulsar passes through the disc. The pressure from the disc will reduce the momentum rate ratio of the pulsar and the stellar wind ($\sim0.01\eta$). For Model C, the disc direction is set as $(i_{\mathrm{d}},\phi_{\mathrm{d}})=(60^{\circ},180^{\circ})$. The disc direction will significantly affect the RM profiles (see Section \ref{201124}). The radio signals travel through different components of the disc when a pulsar passes through the disc near periastron, whcih leads to complicated RM variations (e.g., with some minor structures and multiple reversals).

The DM contributions and the free-free optical depth of Model C (see Fig. \ref{dm}) are also unimportant. Both DM and the free-free optical depth from the disc show the asymmetric profile before and after the periastron.

\section{Clumps in stellar wind and disc}\label{sec:clumps}
The density of the smooth stellar wind and disc is discussed in Section \ref{dmrm}. However, the stellar wind \citep{Moffat1994,Puls2006} or disc \citep{Chernyakova2021} are found to be clumpy from observations. The size of the inhomogeneities is estimated to be $\sim 10^{10}-10^{11}$ cm from the several hours-duration X-ray flares of the Be star binary system 2S 0114+65 \citep{Apparao1991}. The short variability on GeV light curve (between a few minutes to a few days) is also be found from the 2017 \citep{Chernyakova2020} and 2021 periastron passage \citep{Chernyakova2021} of PSR B1259$-$63/LS 2883. The GeV flares are thought to be caused by the bremsstrahlung emission from the clumps in the stellar wind \citep{Chernyakova2020}. The typical size of the clumps can be estimated as
\begin{equation}
    l_{\mathrm{c}}\sim v\Delta t_{\mathrm{f}}=7.2 \times 10^{10} ~\mathrm{cm}\left(\frac{M_{\mathrm{tot}}}{30 ~M_{\odot}}\right)^{1 / 2}\left(\frac{a}{1\times 10^{13} ~\mathrm{cm}}\right)^{-1 / 2}\left(\frac{\Delta t_{\mathrm{f}}}{1 ~\mathrm{h}}\right),
\end{equation}
where $v$ is the Keplerian velocity of the pulsar and $\Delta t_{\mathrm{f}}$ is the durations of the flares. The clumps enhance the density by a factor of $1/f$ compared to the smooth stellar outflows, with $f$ being the filling factor. From observations, the filling factor of the stellar wind is $f\sim0.1$ \citep{Moffat1994,Puls2006}. The inhomogeneities of the density and magnetic field in clumps would lead to stochastic DM and RM variations, which have been detected \citep{Pleunis2021,Mckinven2022}. The minor structures of RM evolution of FRB 20201124A are considered to be caused by the clumps in the disc \citep{Wang2022}. We will give the DM and RM contributions only from the density inhomogeneities of clumps \citep{Melatos1995}.

\begin{figure}
    \centering
    \includegraphics[width = 0.7\textwidth]{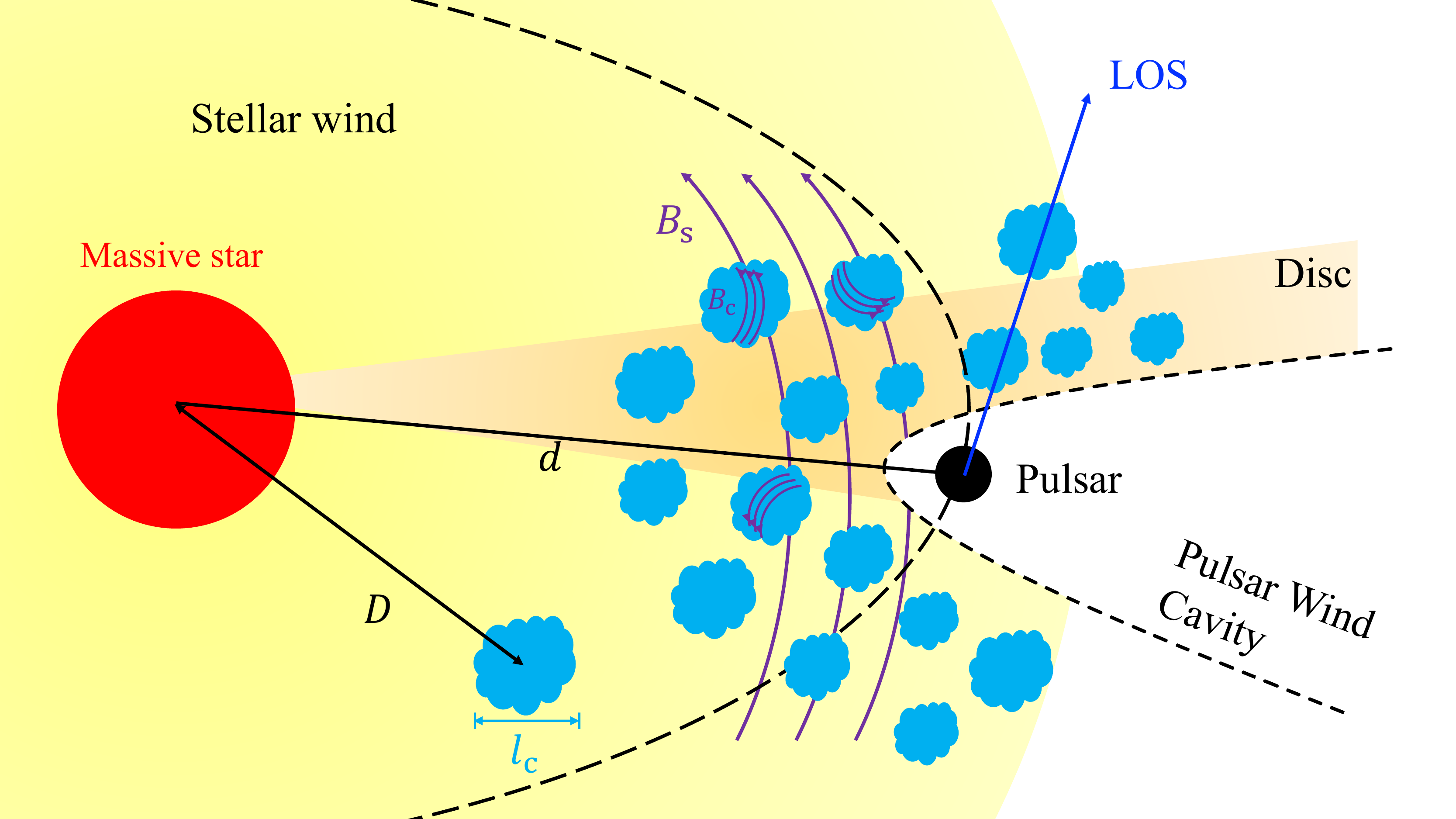}
    \caption{A schematic diagram of interactions between the clumpy stellar wind/disc and the pulsar wind. The inhomogeneities of the density and magnetic field in clumps would lead to stochastic DM and RM variations.}
    \label{clump}
\end{figure}

From our estimation in Section \ref{dmrm} ($P=1600-16000$ d, $M_{\mathrm{tot}}\sim30~M_{\odot}$), the distance from the massive star around periastron is $d\sim10^{14}$ cm. As shown in Fig. \ref{clump}, clumpy stellar wind/disc interacts with the pulsar wind and X-ray/GeV $\gamma$-ray flares generate when clumps enter the LOS. The clump density in stellar wind is $n_{\mathrm{c,w}}=n_{\mathrm{w}}(D)/f$, where $D$ is the distance between clumps and the massive star. For simplicity, we assume the disc clumps are on the mid-plane of the disc. Thus, the clump density in stellar disc is $n_{\mathrm{c,d}}=n_{\mathrm{d}}(D)/f$. The DM from the clumps in stellar wind and disc is
\begin{equation}
    \mathrm{DM_{c,w}} \sim 0.0016 ~\mathrm{pc ~cm^{-3}} \left(\frac{n_{\mathrm{w,0}}}{10^{8}~\mathrm{cm^{-3}}}\right)
    \left(\frac{D}{10^{14}~\mathrm{cm}}\right)^{-2}\left(\frac{f}{0.1}\right)^{-1}
    \left(\frac{l_{\mathrm{c}}}{10^{11}~\mathrm{cm}}\right),
\end{equation}
\begin{equation}
    \mathrm{DM_{c,d}} \sim 0.001 ~\mathrm{pc ~cm^{-3}} \left(\frac{n_{\mathrm{d,0}}}{10^{11}~\mathrm{cm^{-3}}}\right)
    \left(\frac{D}{10^{14}~\mathrm{cm}}\right)^{-3.5}\left(\frac{f}{0.1}\right)^{-1}
    \left(\frac{l_{\mathrm{c}}}{10^{11}~\mathrm{cm}}\right).
\end{equation}
The RM from the radial magnetic field in the stellar wind is
\begin{equation}
    \mathrm{RM_{c,w}} \sim 0.06 ~\mathrm{rad ~m^{-2}} \left(\frac{n_{\mathrm{d,0}}}{10^{11}~\mathrm{cm^{-3}}}\right)\left(\frac{B_0}{1~\mathrm{G}}\right)
    \left(\frac{D}{10^{14}~\mathrm{cm}}\right)^{-4}\left(\frac{f}{0.1}\right)^{-1}
    \left(\frac{l_{\mathrm{c}}}{10^{11}~\mathrm{cm}}\right).
\end{equation}
For a toroidal magnetic field, the RM from the clumpy wind is
\begin{equation}
    \mathrm{RM_{c,w}} \sim 8.8 ~\mathrm{rad ~m^{-2}} \left(\frac{n_{\mathrm{w,0}}}{10^{8}~\mathrm{cm^{-3}}}\right)\left(\frac{B_0}{1~\mathrm{G}}\right)
    \left(\frac{D}{10^{14}~\mathrm{cm}}\right)^{-3}\left(\frac{f}{0.1}\right)^{-1}
    \left(\frac{l_{\mathrm{c}}}{10^{11}~\mathrm{cm}}\right).
\end{equation}
The RM from the clumps on the mid-plane of the disc is
\begin{equation}
    \mathrm{RM_{c,d}} \sim 5.1 ~\mathrm{rad ~m^{-2}} \left(\frac{n_{\mathrm{d,0}}}{10^{11}~\mathrm{cm^{-3}}}\right)\left(\frac{B_0}{1~\mathrm{G}}\right)
    \left(\frac{D}{10^{14}~\mathrm{cm}}\right)^{-4.5}\left(\frac{f}{0.1}\right)^{-1}
    \left(\frac{l_{\mathrm{c}}}{10^{11}~\mathrm{cm}}\right).
\end{equation}
The nearly unchanged DM and small RM fluctuations are consistent with the observations \citep{Pleunis2021,Mckinven2022}. The discussions above are based on the assumption that the inhomogeneities only affect clumps' density. However, the magnetic field in clumps will also contribute to the RM variations although the magnetic field in clumps is still unknown. The sudden RM reversal in a few days through the 2004 periastron passage of PSR B1259$-$63 \citep{Johnston2005} favors a scenario in which the magnetic field reverses in clumps. Whether the RM reversal can be observed on short timescales (i.e., $<1$ day) remains to be tested for FRBs.

{\section{Discussion}\label{discussion}}
{\subsection{DM and RM variations for other repeating FRBs or massive binaries}}
{\subsubsection{DM and RM variations in FRB 20121102A}}

DM and RM variations were first detected in FRB 20121102A. The increasing DM \citep{Hessels2019,Josephy2019,Oostrum2020,LiDi2021} and decreasing RM \citep{Michilli2018,Hilmarsson2021} was found from
long-term observations. The DM and RM variations may be from the wind nebula and the ejecta of a young magnetar born in the merger of a compact binary \citep{Zhao2021a}. A possible period of $\sim160$ d was reported for FRB 20121102A \citep{Rajwade2020,Cruces2021}. Due to lack the study of chromaticity, the origin of periodicity is difficult to determine. Interestingly, its DM has been found to be 552.5 $\pm$ 0.9 pc cm$^{-3}$ from the
recent observations \citep{WP2022}, which has decreased by $\sim10$ pc cm$^{-3}$ from that of \cite{LiDi2021}. If the decrease of DM can be further confirmed by subsequent observations, the DM variations can be well-understood in our model. Due to the difference in pulsar wind cavity size along the LOS at different orbital phases, the evolution of the DM before and after periastron can be asymmetric (see Fig. \ref{dm}).

The extreme RM ($\sim10^{5}$ rad m$^{-2}$, \citealt{Michilli2018,Hilmarsson2021}) of FRB 20121102A may be from the compact PRSs \citep{Margalit2018,Zhao2021b,Yang2022a,Sridhar2022}. Therefore, there may be two possible RM evolution trends of FRB 20121102A afterward. If the RM from the binary (RM$_{\text{b}}$, e.g., from the stellar wind/disc or clumps) is much smaller than that from PRS, RM will just decrease. However, if RM$_{\text{b}} \simeq \mathrm{RM_{PRS}}$, the irregular RM variations or reversals will be detected.

\subsubsection{RM variations and reversals in FRB 20190520B}\label{190520}

The RM variations and reversals of FRB 20190520B \citep{Anna-Thomas2022,Dai2022} have been discussed in \cite{Wang2022} and \cite{Dai2022} using the binary model. \cite{Dai2022} consider that the FRB source orbits around a magnetized companion (e.g., a massive black hole or a massive star). The large-scale magnetic field is radial. The pulsar will travel through the magnetic field reversal region when the pulsar orbit is edge-on with respect to the magnetic axis. Here, we will give a general description of the case of the radial magnetic field. The orbital-dependent normalized RM variations of different magnetic axis inclination angles are shown in Fig. \ref{rm-phi}(a), and other parameters are the same as Model A. We can see that the RM does not reverse, except if the magnetic axis inclination angle is close to $90 ^\circ$.

Another explanation introduces the toroidal magnetic field in the decretion disc of a Be star \citep{Wang2022}. In this work, some complex RM variations caused by the additional pressure and the clumps from the disc are also considered, which can be checked by future observations. In addition to the above two cases, the toroidal magnetic field in the stellar wind is also possible.

\subsubsection{Complicated RM variations in FRB 20201124A and PSR B1259$-$63}\label{201124}

\begin{figure}
    \centering
    \includegraphics[width = 1\textwidth]{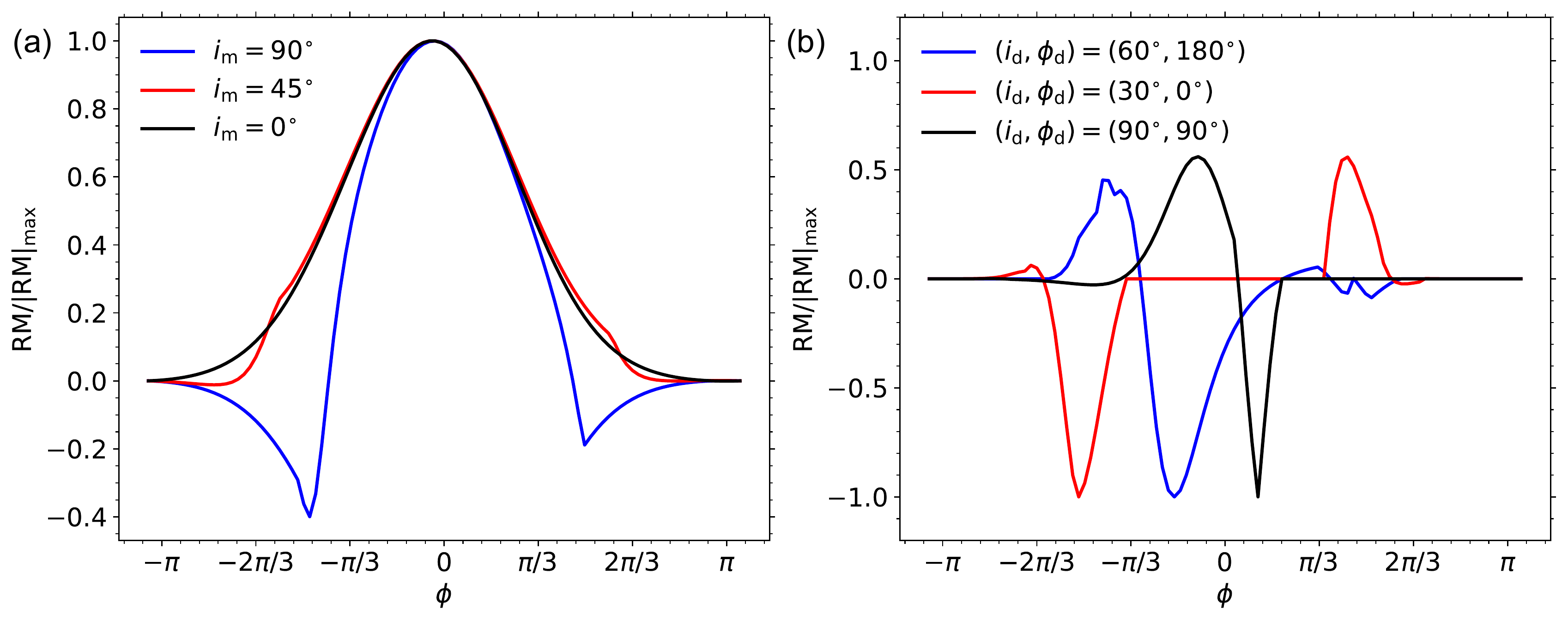}
    \caption{The orbital-dependent normalized RM of different magnetic axis inclination angles (panel a) and disc directions (panel b). (a) The wind magnetic field is radial and other parameters are the same as Model A. The RM does not reverse, except if the magnetic axis inclination angle is close to $90 ^\circ$. (b) The disc magnetic field is toroidal and other parameters are the same as Model C. RMs are complicated and profiles are completely different depending on the disc directions.}
    \label{rm-phi}
\end{figure}

The irregular RM variations and multiple reversals in one cycle have been observed for FRB 20201124A \citep{Xu2022} and PSR B1259$-$63 \citep{Johnston1996,Connors2002,Johnston2005}. For PSR B1259$-$63, RMs can even be different between two periastron passages \citep{Johnston1996,Connors2002,Johnston2005}. The complexity of RM variations may not be explained by uncertainties in orbital parameters and observation angles because these parameters are well constrained by pulsar timing and very long baseline interferometric observations \citep{Shannon2014,Miller-Jones2018}. There are several possible reasons for the complexity of RM variations. First, some random variations or even reversals are caused by the clumps in the stellar wind/disc \citep{Wang2022}. Second, RM contributed by the stellar wind and disc (or/and the radial and toroidal magnetic field) both exist, which results in a multimodal and multiple reversal profile of the RM evolution. Third, the disc direction will have a great impact on the evolution of RM. The complicated RM profiles of different disc directions are shown in Fig. \ref{rm-phi}(b), and other parameters are the same as Model C. Fourth, the local disc density and magnetic field may be different when the pulsar enters the disc \citep{Johnston2005}, which may be the reason why RMs are different between two periastron passages \citep{Johnston1996,Connors2002,Johnston2005}.

The complicated RM variations caused by the disc can be distinguished by their behavior near periastron (see Fig. \ref{rm-phi}(b)). For PSR B1259$-$63, the pulsar is eclipsed by the dense stellar outflows around periastron. 
Luckily, the eclipsing seems to be not important for FRB 20201124A \citep{Xu2022,Wang2022}. The interaction between the pulsar and the disc makes RM$_{\mathrm{disc}}$ unpredictable. However, from the observations of 
different periastron passages of PSR B1259$-$63 \citep{Johnston1996,Connors2002,Johnston2005}, the evolutionary trend of the RM is similar after the pulsar leaves the disc. Different from the disc, stellar winds are stable on 
long-time scales. The eclipse near periastron is mainly caused by the stellar wind for PSR B1259$-$63, which will not be significantly different during different periastron passages. Thus, constraints on orbital parameters and complex 
environments can still be given via the long-term monitoring of FRB 20201124A.

{\subsection{Compact persistent radio sources}}

FRB 20121102A \citep{Chatterjee2017} and FRB 20190520B \citep{Niu2022} are associated with compact luminous ($\nu L_{\nu} \sim 10^{39}$ erg s$^{-1}$) PRSs. The compact PRS may be from a magnetar wind nebula \citep{Margalit2018,Zhao2021b}, a magnetized plasma screen \citep{Yang2022a} or ultraluminous X-ray hypernebulae \citep{Sridhar2022}. All the above models consider PRSs to be associated with large RMs ($\sim10^{4}-10^{5}$ rad m$^{-2}$) and that the age of center engine is very young (e.g., 14-22 yr in \citealt{Zhao2021b} and 10 yr in \citealt{Sridhar2022}). There is no evidence for the PRS of FRB 20180916B \citep{Marcote2020}.% luminosity of the PRS associated with FRB 20180916B is too low ($\nu L_{\nu} < 10^{36}$ erg s$^{-1}$) to be related to FRBs \citep{Marcote2020}. }

Considering that the binary model of \cite{Sridhar2022} has evolved as a common-envelope precursor, we will not discuss this model here. Besides, there are two possible origins of luminous PRS in massive binaries: the scaled-down pulsar wind nebulae (PWNe, \citealt{Margalit2018,Zhao2021b}) in HMGBs or the companion wind as the plasma screen \citep{Yang2022a}. The structures of the compact scaled-down PWNe have been found in HMGBs \citep{Neronov2007,Zdziarski2010}. As discussed in \cite{Zhao2021b}, the interior magnetic energy of young magnetars will be injected into the surroundings (here being the scaled-down PWNe or the companion wind) via energetic magnetar flares and synchrotron radiation of relativistic electrons produce luminous PRSs.

\section{Summary}\label{conclusion}
Recently, the RM variations or even reversals have been observed for FRB 20121102A \citep{Michilli2018}, FRB 20201124A \citep{Xu2022}, FRB 20190520B \citep{Anna-Thomas2022,Dai2022} and FRB 20180916B \citep{Pleunis2021,Mckinven2022}, which are similar to the RM changes of PSR B1259$-$63/LS 2883. Some binary models have been proposed to explain the RM variations and reversals \citep{Dai2022,Wang2022}. Compared to previous work, we give a more general description of the DM, RM and free-free absorption process from a massive binary system. We study the contributions of RM, DM and free-free absorbtion process from the stellar wind and disc, and fit the RM variations of FRB 20180916B. This paper presents the effects of orbital geometry, magnetic field structure, the pulsar wind cavity, observation (magnetic axis, disc) orientation and the clumpy stellar wind/disc on DM/RM variations, and these differences can be tested by subsequent observations. Our conclusions are summarized as follows.

\begin{itemize}
    \item For the case of the radial magnetic field in the stellar wind, RM will not reverse except if the magnetic axis inclination angle is close to $90 ^\circ$. However, for the toroidal magnetic field in the stellar wind, RM will reverse before and after the SUPC.
    \item For the case of the toroidal disc magnetic field domain, the RM variations become complicated when the pulsar passes through the inclined disc twice, i.e., before and after periastron. The RM from the disc may have a multimodal and multiple reversal profile because the radio signals travel through different components of the disc during periastron passage. What's more, the interaction between the pulsar and the disc makes RM different between two periastron passages \citep{Johnston1996,Connors2002,Johnston2005}.
    \item In this work, we assume the 16.35-day period and the chromaticity are from a slowly rotating or freely precessing magnetar \citep{LDZ2021}. The secular RM variations of FRB 20180916B are raised from periastron passage of a pulsar with the orbital period of $\sim1600-16000$ d. The DM and free-free absorption contributed by the stellar wind and disc are negligible.
    \item When the clumps in stellar wind or disc interact with FRBs, the inhomogeneities of the density and magnetic field in clumps would lead to stochastic RM variations or even reversals.
\end{itemize}

\section*{acknowledgements}
We thank the anonymous referee for helpful comments, and Yuan-Pei Yang and A-Ming Chen for helpful discussions. This work was supported by the National Natural Science Foundation of China (grant Nos. 12273009, U1831207 and 11833003), the
National Key Research and Development Program of China
(grant No. 2017YFA0402600), the National SKA Program of
China (grant Nos. 2020SKA0120300 and 2022SKA0130100), and the China Manned Spaced Project (CMS-CSST-2021-A12).

\bibliographystyle{aasjournal}
\bibliography{ms}

\begin{table}[]
    \centering
    \caption{Meanings and Typical Values of Physical Parameters in the model.}
    \label{tab:my_label}
	\begin{threeparttable}
	    \begin{tabular}{clccc}
	  \hline
      & Parameter & Symbol & Value & Reference \\
      \hline
      & Mass of star &  $M_{\star}$ & 30 $M_{\odot}$ &   (1)\\
      Star & Radius of star & $R_{\star}$ &  $\sim10R_{\odot}$ &   (1,2)\\
      & Effective temperature of star & $T_{\star}$ & $3\times10^4$ K & (1)\\
      \hline
      & Mass loss rate & $\dot{M}$ &  $10^{-11}-10^{-8} M_{\odot}~ \mathrm{yr}^{-1}$ & (3,4) \\
      Stellar wind & Wind velocity & $v_{\mathrm{w}}$ & $3 \times 10^{8} \mathrm{~cm} \mathrm{~s}^{-1}$ & (3,4)\\
      & Wind temperature slope & $\beta_{\mathrm{w}}$ & 2/3 & (5,6)\\
      \hline
      & Disc density near stellar surface & $\rho_{\mathrm{d}, 0}$ & 10$^{-13}- 10^{-10}$ g cm$^{-3}$ & (7)\\
      & Disc density slope & $\beta_{\mathrm{d}}$ & 3.5 & (7,8)\\
      Stellar disc  & Inclination angle of the disc normal vector & $i_{\mathrm{d}}$ & $60^{\circ}$ & (9,10)\\
      & True anomaly angle of the disc normal vector & $\phi_{\mathrm{d}}$ & $180^{\circ}$ & \\
      & Half-opening angle & $\Delta \theta$ & $15^{\circ}$ & (11)\\
      \hline
      Pulsar & Mass of pulsar & $m$ & 1.4 $M_{\odot}$ &   \\
      &  Spin-down luminosity & $L_{\mathrm{sd}}$ & $10^{35}-10^{36}$ erg s$^{-1}$ & (12,13)\\
      \hline
    \end{tabular}
    \begin{tablenotes}
	   % \footnotesize
	    \item References: (1) \cite{Negueruela2011}; (2) \cite{Casares2005}; (3) \cite{Snow1981}; (4) \cite{Krticka2014}; (5) \cite{Kochanek1993}; (6) \cite{Bogomazov2005}; (7) \cite{Rivinius2013}; (8) \cite{Carciofi2006}; (9) \cite{Chen2021}; (10) \cite{Martin2009};
	    (11) \cite{Waters1986}; (12) \cite{Manchester1995}; (13) \cite{Camilo2009};
    \end{tablenotes}
	\end{threeparttable}
\end{table}

\begin{table}
    %\resizebox{\textwidth}{!}{
    \centering
        \caption{The model parameters of the RM fitting results for FRB 20180916B.}
    \label{parameters}
    \begin{tabular}{clccccc}
    \hline
            & Parameter & Symbol &  Unit & Model A & Model B & Model C \\
    \hline
            & Period    & $P$    &  d & 1600 & 2000 & 16000\\
    Orbital & Eccentricity & $e$ &    & 0.5 & 0.5 & 0.86 \\
            & Periastron time & $T_0$ & MJD & 59650 & 59900 & 59700 \\
    \hline
    Observational & Inclination angle of observers &  $i_{\mathrm{o}}$  & & $26^{\circ}$ & $154^{\circ}$ & $45^{\circ}$ \\
    angles & True anomaly angle of observers & $\phi_{\mathrm{o}}$ & & $132^{\circ}$ & $132^{\circ}$ & $132^{\circ}$ \\
    \hline
    Pulsar & Spin-down luminosity & $L_{\mathrm{sd}}$ &  erg s$^{-1}$ & $10^{36}$  & $10^{35}$  & $10^{36}$ \\
    \hline
     & Mass loss rate & $\dot{M}$ &  $ M_{\odot}~ \mathrm{yr}^{-1}$ &  $10^{-8}$ &$10^{-9}$ &$10^{-8}$\\
                     & Surface magnetic field & $B_0$ & G & 5 & 2 & 1 \\
    Stellar outflows & Wind magnetic field slope & $\alpha$ & & 2 & 1 & 2\\
                     & Disc magnetic field slope & $\alpha$ & &  &  & 1\\
                     & Disc density near stellar surface & $\rho_{\mathrm{d}, 0}$ &  g cm$^{-3}$ & & & $1.2\times10^{-13}$\\
    \hline
    \end{tabular}
    %}
\end{table}

\end{document}